\documentstyle[11pt]{article}

\newcommand{\beq}{\begin{equation}}
\newcommand{\eeq}{\end{equation}}
\newcommand{\bea}{\begin{eqnarray}}
\newcommand{\eea}{\end{eqnarray}}

\newcommand{\chii}{\raise.5ex\hbox{$\chi$}}

\newcommand{\R}{I \! \! R}

\newcommand{\noi}{\vspace{12pt}\noindent}

\newcommand{\ie}{{${ i.e.\ }$}}

\newcommand{\ea}{{${ et~al.\ }$}}

\newcommand{\e}[1]{{(\ref{#1})}}
\newcommand{\eq}[1]{{eq.\ (\ref{#1})}}
\newcommand{\es}[2]{{(\ref{#1}) and (\ref{#2})}}
\newcommand{\eqs}[2]{{eqs.\ (\ref{#1}) and (\ref{#2})}}
\newcommand{\Ref}[1]{{Ref.~\cite{#1}}}

\newcommand{\lpart}{\raise.3ex\hbox{$\stackrel{\leftarrow}{\partial}$}\!}
\newcommand{\rpart}{\raise.3ex\hbox{$\stackrel{\rightarrow}{\partial}$}}
\newcommand{\ldr}{\raise.3ex\hbox{$\stackrel{\leftarrow}{\delta}$}}
\newcommand{\deder}[1]{{ 
 {\stackrel{\raise.1ex\hbox{$\leftarrow$}}{\delta}   } 
\over {   \delta {#1}}  }}
\newcommand{\dedel}[1]{{ 
 {\stackrel{\lower.3ex \hbox{$\rightarrow$}}{\delta}   }
 \over {   \delta {#1}}  }}

\newcommand{\papar}[1]{{ 
 {\stackrel{\raise.1ex\hbox{$\leftarrow$}}{\partial}   } 
\over {   \partial {#1}}  }}
\newcommand{\papal}[1]{{ 
 {\stackrel{\lower.3ex \hbox{$\rightarrow$}}{\partial}   }
 \over {   \partial {#1}}  }}

\newcommand{\Hf}{{1 \over 2}}

\newcommand{\Ih}{{i \over \hbar}}

\newcommand{\V}{\vartheta}

\newcommand{\norma}{\frac{1}{N}}
\newcommand{\twonorma}{\frac{2}{N}}
\newcommand{\cech}[1]{\breve{#1}}

\begin{document}
\thispagestyle{empty}
\vspace{3cm}
\title{\Large{\bf Hamiltonian Superfield Formalism\\
  with $N$ Supercharges}}
\vspace{2cm}
\author{
{\sc I.A.~Batalin}$^{\,\dag}$\\
Lebedev Physics Institute\\ 53 Leninisky Prospect\\
Moscow 119991, Russia,\\
~\\and\\~\\
{\sc K.~Bering}$^{\,\ddag}$\\
University of Illinois at Chicago\\
845 W Taylor St.\ (M/C~273)\\
Chicago, IL~60607-7059, USA\\
\vspace{3cm}}
\maketitle
\begin{abstract}
An action principle that applies uniformly to any number $N$ of 
supercharges is proposed. We perform the reduction to the $N=0$ 
partition function by integrating out superpartner fields. 
As a new feature for theories of extended supersymmetry, 
the canonical Pfaffian measure factor is a result of a 
Gaussian integration over a superpartner.
This is mediated through an explicit choice of direction $n^a$
in the $\theta$-space, which the physical sector does not depend on.  
Also, we re-interpret the metric $g^{ab}$ in the Susy algebra 
$[D^a,D^b] \sim g^{ab}\partial_t$ as a symplectic structure 
on the fermionic $\theta$-space. This leads to a superfield 
formulation with a general covariant $\theta$-space sector. 
\end{abstract}

\begin{quote}

\vfill 
PACS number(s): 04.20.Fy, 04.60.Gw, 11.15.--q, 12.60.Jv. 
\rightline{\tt arXiv:hep-th/0401169} 
\hrule width 5.cm \vskip 2.mm 
$^{\dag}${\small \noindent E-mail:~{\tt batalin@lpi.ru}} \\
$^{\ddag}${\small \noindent E-mail:~{\tt bering@uic.edu}}

\end{quote}

\newpage

\setcounter{equation}{0}
\section{Introduction: Review of $N=1$}

\noi
A few years ago, we developed a $N\!=\!1$ superfield formulation of 
Hamiltonian field theories \cite{BBD1,BBD2}, where all the fields $z^A_0$ 
are replaced by superfields
\beq
z^A~=~z^A_0+ \theta z^A_1~.
\eeq
The basic idea is that $\theta$-translations should encode the 
BRST symmetry, so that the superpartners  
\beq
z_1^A ~=~ \{\Omega(z_0),z^A_0\}_{PB}
\eeq
are the corresponding BRST-transformed fields. Furthermore, the 
supersymmetry should be implemented as a square root of time translations
\beq
  D^2~=~{\partial \over {\partial t}}~.
\eeq 
As the superderivative is a combination of $\theta$ and $t$ derivatives,
\beq
  D~=~{\partial \over {\partial \theta}} 
+ \theta {\partial \over {\partial t}}~,
\eeq
one should introduce a matching supercharge, which is a combination 
of the BRST charge and the Hamiltonian\footnote{
{}For comparison, the signs of $Q$ and $\Omega$ are the opposite 
of the conventions used in \Ref{BBD1,BBD2}.}
\beq
 Q~=~\Omega - \theta H~.
\label{q1}
\eeq
It is natural to postulate the following ``superequations'' of motion 
\beq
 D F~=~\{Q , F\}_{PB}+D_{{\rm expl}}F ~
\label{qeom1}
\eeq
for any quantity $F=F(z(t,\theta);t,\theta)$, where ``expl'' denotes 
explicit differentiation. Applying the equations of motion twice on 
a field $z^A$, we get the equations for time evolution
\beq
   \dot{z}^A~=~-\{ H , z^A \}_{PB}~,
\eeq 
if we let the Hamiltonian be
\beq
  - H ~=~ \Hf \{ Q,Q \}_{PB}+ D_{{\rm expl}} Q~.
\label{ham1}
\eeq
Note that $H$ is expressed entirely in terms of the supercharge $Q$.
Similarly, \eq{q1} can be thought of as a definition of $\Omega$
in terms of $Q$. At this point, the definition \e{ham1} appears to be
a very restrictive choice of Hamiltonian, but this is not so. 
If we for instance consider $\Omega$ to be nilpotent and without 
explicit $\theta$-dependence (as would normally be the case), 
then {\em any} BRST-invariant Hamiltonian can be written in the form 
\e{ham1}! For further details, including the necessary integrability 
conditions, we refer to the \Ref{BBD1,BBD2}. Lagrangian and related 
superfield formulations have been discussed in \Ref{BBD1,superBV} and
\Ref{other}, respectively.

\subsection{Path Integral}

\noi
The $N\!=\!1$ operator and path integral formalism were worked out 
in \Ref{BBD1,BBD2}. Here we review the $N\!=\!1$ path integral 
construction, as it serves as an important prototype for further developments. 
Assume in the following that the supercharge $Q$ has no explicit time 
dependence. The action reads
\bea
S&=&\int dt~d\theta~
\left[z^A \bar{\omega}_{AB}(z)~Dz^B(-1)^{\epsilon_B}+Q(z)\right] 
\label{s1} \\ 
&=&\int dt\left[ z_0^A \bar{\omega}_{AB}(z_0)~ \dot{z}_0^B
+ \Hf z^A_1 \omega_{AB}(z_0)~  z^B_1(-1)^{\epsilon_B} 
+ (z^A_1\partial_A + \partial_{\theta}^{{\rm expl}}) Q(z_0) \right]~,
\label{sc1}
\eea
where we in the last expression have written the action out in components.
({}For details concerning the symplectic $2$-tensor $\omega_{AB}$, see 
Section~\ref{secsympz} and \eq{omegabar}.) Note that the superpartners 
$z^A_1$ only appear up to the Gaussian order in the action \e{sc1}.
Hence it is possible to integrate them out of the path integral.
This reduces in a very direct way the $N\!=\!1$ path integral 
to the original $N\!=\!0$ partition function:
\beq
  {\cal Z}~=~ \int [dz_0][dz_1] \exp\left[\Ih S \right]
  ~=~\int [dz_0] ~{\rm Pf}(\omega_{AB}(z_0))~  \exp\left[\Ih \int dt\left( 
z_0^A \bar{\omega}_{AB}(z_0)~\dot{z}_0^B-H(z_0)\right)\right]~.
\label{pathint1}
\eeq
Remarkably the Gaussian integration has just the right impact on 
both the classical and the quantum part of the partition function: 
 
\begin{enumerate} 
\item
Completing the square generates a shift-term in the action, which
restores the Hamiltonian to the form \e{ham1}. 
\item
The Gaussian integration produces the canonical Pfaffian measure factor 
of the the original $N\!=\!0$ formalism. The factor is needed 
(at the naive and purely formal level) to maintain a reparametrization 
invariant path integral. 
\end{enumerate}
The simplicity of the above reduction \e{pathint1} from $N\!=\!1$ 
to $N\!=\!0$ suggests that it should serve as a cornerstone 
for further theoretical developments of the superfield 
formalism.

\subsection{The Plan of the Paper}

\noi
The purpose of the paper is twofold:
\begin{enumerate}
\item
We would like to generalize to higher supersymmetries \cite{Hull}, first 
and foremost to the $N\!=\!2$ case (see Section~\ref{secn2} below). Such 
a situation arises in theories with both a BRST and an anti-BRST symmetry.
The Susy algebra and the equations of motion are easy to generalize
\cite{BD}. The main hurdle is the formulation of an action principle 
that gives rise to a correct path integral. As a minimum requirement, 
any proposals for extended supersymmetry should reduce to the original 
$N\!=\!0$ partition function {\em via} the above $N\!=\!1$ path integral 
\e{pathint1}. Previous works \cite{BD,Gozzi} do not meet this test. 
{}For instance, the path integral of Gozzi \ea \cite{Gozzi} is confined 
to the classical trajectories. In their approach, the rigid $N\!=\!2$ 
geometry has completely ironed out quantum fluctuations. The proposal 
of \Ref{BD} has correct quantum behavior in the original $z_0$-sector, 
but their superpartner $z_1$ is constrained, in contrast to the 
superpartner $z_1$ in the $N\!=\!1$ action \e{sc1}. Furthermore, the 
quantum measure factor of \Ref{BD} is generated with the help of a 
vielbein $h_A{}^B$. We shall here give a new proposal that remedies 
this with the caveat, that we have explicitly selected a 
$\theta$-direction $n^a$. However, the path integral does not depend on $n^a$.
\item
{}For $N>2$ an additional challenge arises, as flat coordinate systems 
for the $\theta$-space are no longer protected (by the requirement of
no external fermionic constants). Hence, it is of interest to develop a 
general covariant theory for the $\theta$-space (see Section~\ref{genn}).
\end{enumerate}

\setcounter{equation}{0}
\section{$N=2$ Revisited}
\label{secn2}

\noi
The $N=2$ case is physically motivated by Hamiltonian $Sp(2)$-symmetric 
theories \cite{BLT}. Such theories are endowed with a BRST and an anti-BRST 
charge $\Omega^a$, $a=1,2$, of ghost number $\pm 1$, respectively.
The theories are invariant under rotations $\Omega^a=\Lambda^a{}_b~\Omega'^b$
with $2 \times 2$ matrices  $\Lambda^a{}_b \in Sp(2) \cong SL(2,\R)$.
The idea is now to implement geometrically the two fermionic 
BRST/anti-BRST symmetries by introducing two fermionic parameters 
$\theta_a$, $a=1,2$. The $N=2$ superfield 
\beq
 z^A~=~z^A_0+ \theta_a z^{aA}+\theta^2 z_3^A~,~~~~~~~~~~~~~~~~~~~~~~~~
\theta^2 ~:=~ \Hf \epsilon^{ab}\theta_a \theta_b~=~\theta_1 \theta_2~,
\eeq
has four component fields, $z^A_0$, $z^{1A}$, $z^{2A}$ and $z^A_3$. Of these 
four fields, the three superpartners are BRST/anti-BRST transformed fields, 
\bea
z^{aA}&=& \{\Omega^a(z_0),z^A_0\}_{PB}~,~~~~~~~~~~~~a=1,2~, \\
z^A_3 &=& \Hf \epsilon_{ab} \left(
\{\Omega^a(z_0),\{\Omega^b(z_0),z^A_0\}_{PB}\}_{PB}
+\{\partial_{{\rm expl}}^a \Omega^b(z_0),z^A_0\}_{PB}\right) ~.
\eea
The three ``BRST/anti-BRST transformation laws'' are implemented 
via a suitable choice of equations of motion (see Section~\ref{seclaseq}).

\subsection{The Metric $g^{ab}$}
\label{metric2}

\noi 
We can now move around along two linearly independent directions in the 
fermionic $\theta$-plane. Therefore, we need to introduce two 
superderivatives $D^a$, $a=1,2$, such that the Susy algebra 
{\em inevitably} acquires a symmetric $2 \times 2$ metric\footnote{
{}For comparison, the reference \Ref{BD} has a metric 
$g^{ab}_{{\rm there}}=\Hf g^{ab}_{{\rm here}}$.} $g^{ab}$: 
\beq
 [D^a,D^b]~\equiv~D^{\{a}D^{b\}}~=~g^{ab} {\partial \over {\partial t}}
~,~~~~~~~~~~~~~~~~~~~~~g^{ab}~=~g^{ba}~.\label{susyalg2}
\eeq
The appearance of a symmetric metric $g^{ab}$ is perhaps the single most 
important new feature for the $N\!=\!2$ case, so let us investigate it in
further detail. It has effectively two degrees of freedom, as an overall 
normalization just rescales the time variable $t$. It transforms as a 
tensor $g^{ad}=\partial^a\theta'_b~ g'^{bc}~\partial^d\theta'_c$ under 
reparametrization of the fermionic coordinates $\theta_a \to \theta'_b$.
As we ignore a less attractive possibility\footnote{
This possibility is covered in complete generality in Section~\ref{genn}.
However, there is no physical motivation to introduce external 
fermionic parameters in a theory.} of introducing fermionic constants, 
the most general reparametrization is of the form 
$\theta'_b=\theta_a \Lambda^a{}_b$, where $\Lambda^a{}_b \in GL(2,\R)$ is
a constant bosonic matrix with ${\rm det}(\Lambda^a{}_b) \neq 0$, \ie the
geometry is completely rigid. Arguments along similar lines combined with 
the fact that $g^{ab}$ should be cohomologically closed (see 
Section~\ref{symplectheta}), show that the metric $g^{ab}$ does not 
depend on $\theta_a$.

\noi
The origin of the metric tensor $g^{ab}$ can be related to a real 
$2 \times 2$ matrix $G^a{}_b$ for the ghost number operator $G$:
\beq
      \{G,\Omega^a\}_{PB}~=~G^a{}_b~\Omega^b~.
\eeq
This matrix $G^a{}_b$ also has two degrees of freedom, because its 
determinant and trace are fixed from the onset:
\beq
{\rm det}(G^a{}_b)~=~{\rm det}(\sigma_3)~=~-1~,~~~~~~~~~~~~~~~~~~
{\rm tr}(G^a{}_b)~=~{\rm tr}(\sigma_3)~=~0~,
\eeq
where $\sigma_3$ is the 3rd Pauli matrix.
One may identify
\beq
\sqrt{-g} g^{ac}~\equiv~G^a{}_b~\epsilon^{bc}~,
~~~~~~~~~~~~~~~~~~~~~~~g~:=~{\rm det}(g_{ab})~<~0~,
\eeq
where there is included a determinant factor $\sqrt{-g}$ on the lhs.\ 
to absorb an inessential overall normalization.  The symmetry of 
$g^{ab}$ is a consequence of $G^a{}_b$ being traceless. 
As the metric $g^{ab}$ should be real, it acquires an indefinite 
$(1,1)$ signature (see Section~\ref{symplectheta}). An indefinite
metric does not pose a fundamental challenge to our construction, 
but it is however a technical nuisance, and for simplicity, we assume
from now on that the metric $g^{ab}$ is positive definite.

\subsection{Equations of Motion}

\noi
In general, we use the metric $g^{ab}$ to raise and lower $Sp(2)$ indices. 
{}For instance, $\theta^a:=g^{ab}\theta_b$. As the superderivatives are 
realized as
\beq
 D^a~=~{\partial \over {\partial \theta_a}}+ 
\Hf \theta^a {\partial \over {\partial t}}~,
\eeq
one should introduce matching supercharges\footnote{
The supercharges $Q^a$ could in principle have both explicit $t$
and explicit $\theta$ dependence. However, whenever we work in a path 
integral formalism, we assume that the $Q^a$ contain no explicit time 
dependence.}
\beq
 Q^a~=~\Omega^a - \Hf \theta^a H~,
\eeq
and impose the following equations of motion
\beq
 D^a F~=~\{Q^a , F\}_{PB}+ D_{{\rm expl}}^a F ~.
\label{qeom2}
\eeq
Applying the equations of motion twice on a field $z^A$, and then 
symmetrizing, we get the equations for time evolution
\beq
   \dot{z}^A~=~-\{ H , z^A \}_{PB}~,
\eeq 
if we let the Hamiltonian be
\beq
 - H ~=~ \Hf g_{ab}\left( \{ Q^a,Q^b \}_{PB}
+ D^{\{a}_{{\rm expl}} Q^{b\}} \right)~.
\label{ham2}
\eeq

\subsection{Two Sectors: Tilde and Check}

\noi
It turns out that the equations of motion \eq{qeom2} in their present 
formulation are not directly applicable for action and path integral 
building. Instead, one can give an equivalent, more tractable
formulation \cite{BD}. {}First, introduce a Hodge-dual
\beq
\tilde{\theta}^a~:=~\frac{\epsilon^{ab}}{\sqrt{g}} ~\theta_{b}
\eeq
to $\theta_a$. (See the Appendices for further details.)
Now build two bosonic supercharges
\bea
\tilde{Q}&:=&\tilde{\theta}_a Q^a 
~=~\tilde{\theta}_a \Omega^a + \delta^2(\theta)~H 
\label{tildeq2} \\
\cech{Q}&:=&\theta_a Q^a~=~\theta_a \Omega^a~.
\label{cechq2}
\eea
and two bosonic superderivatives\footnote{$\cech{D}$ and 
$\cech{Q}$ were denoted $D$ and $Q$, respectively, in \Ref{BD}.}
\bea
\tilde{D}&:=&\tilde{\theta}_a D^a 
~=~\tilde{\theta}_a \partial^a - \delta^2(\theta)~\partial_t 
 \label{tilded2} \\
\cech{D}&:=&\theta_a D^a~=~\theta_a \partial^a~.
\label{cechd2}
\eea
One may show (see Section~\ref{twosec}-\ref{cechsec}) that the 
equations of motion \eq{qeom2} are equivalent to the following set 
of equations:
\bea
\tilde{D}F&=&\{\tilde{Q},F\}_{PB}+ \tilde{D}_{{\rm expl}} F 
\label{tildeeom2} \\
\cech{D}F&=&\{\cech{Q},F\}_{PB}+ \cech{D}_{{\rm expl}} F~,
\label{cecheom2}
\eea
provided the pertinent integrability conditions are satisfied.

\subsection{New Action}

\noi 
Our new action proposal consists of three parts:
\bea
  S&=&\tilde{S}[z]+\cech{S}[z,w]+S_n[w,\pi]~, \label{ss2} \\
\tilde{S}[z]&=&-\int dt \sqrt{g} ~d^2\theta
\left[ z^A \bar{\omega}_{AB}(z)~\tilde{D}z^B +\tilde{Q}(z)\right]~, \\
\cech{S}[z,w]&=&\int dt \sqrt{g} ~d^2\theta~w^A 
\left[  \omega_{AB}(z)~\cech{D}z^B + \partial_A\cech{Q}(z)\right]~, \\
S_n[w,\pi]&=&\int dt \sqrt{g} ~d^2\theta~w^A(\theta)~\pi_A(n^a \theta_a)~.
\eea
Here $z^A$ and $w^A$ are $N\!=\!2$ superfields, while $\pi_A$ is a 
$N\!=\!1$ auxiliary superfield. This yields $2^2+2^2+2^1=10$ components 
for each $A=1,\ldots, 2M$. All three superfields carry the same 
Grassmann-parity $\epsilon_A$. Alternatively, the superfield $w^A$ may 
be viewed as a ``collective field'' or ``shift field'',
\beq
\cech{S}[z,w]~=~
\int dt \sqrt{g} ~d^2\theta\left[w^A \omega_{AB}(z)~\cech{D}z^B 
+\cech{Q}(z\!+\!w)-\cech{Q}(z)+{\cal O}(w^2)\right]~.
\eeq
The main new ingredient is provided by a gauge-fixing real unit-vector 
$n^a$, which satisfies 
\beq
n^a g_{ab} n^b~=~1~.
\eeq
It represents an explicit choice of direction 
$\theta_{\parallel}:=n^a\theta_a$ in the $2$-dimensional $\theta$-plane. 

\noi
We claim that the variation of the action yields the equations of motion
\bea
  \tilde{D}z^A&=&\{\tilde{Q},z^A\}_{PB}~, \label{tilde2eom} \\
   \cech{D}z^A&=&\{\cech{Q},z^A\}_{PB}~,  \label{cech2eom} \\
      w^A &=& 0~,  \label{w2eom} \\
 \pi_A &=& 0   \label{pi2eom}~.
\eea
Parts of this are proved below. It follows immediately that the 
variation of the tilde part $\tilde{S}$  wrt.\ the superfield $z^A$ 
produces the tilde equation of motion \e{tilde2eom}, but there could 
potentially be ``reaction force'' contributions from the second sector 
$\cech{S}$. This is prohibited, as we will see, because the reaction 
terms vanishes, mediated by a marvelous compatibility between the two 
sectors. In particular, we will see that the full $w^A$ superfield are 
annihilated on-shell: $w^A \cong 0$. Similarly, variation of the second 
part of the action $\cech{S}$ wrt.\ the superfield $w^A$ produces the 
second equation \e{cech2eom}. Again possible interfering contributions 
from the third sector $S_n$ are avoided, although the actual details in 
this case are less important, as we are mostly interested in the $z$-sector. 

\noi
The action reads in components
\bea
\tilde{S}[z]&=&\int dt \left[ z_0^A \bar{\omega}_{AB}(z_0)~\dot{z}_0^B \right.
+ \Hf  z^{aA}g_{ab} \omega_{AB}(z_0)~ z^{bB}(-1)^{\epsilon_B} 
\nonumber \\
&&+\left. g_{ab}(z^{aA}\partial_A+\partial^a_{{\rm expl}}) Q^b(z_0)\right]~,
\label{s1c2} \\
\cech{S}[z,w]&=& \int dt~\sqrt{g} \left[ \epsilon_{ab} w^{aA}
[\omega_{AB}(z_0)~z^{bB} 
(-1)^{\epsilon_B} + \partial_A Q^{b}(z_0) ]\right. \nonumber \\
&& + 2  w^A_0 ~  \omega_{AB}(z_0)~z^B_3   
- (-1)^{\epsilon_A} \epsilon_{ab}  z^{aA} z^{bB}
\partial_B \omega_{AC}(z_0)~ w^C_0  \nonumber \\
&& + \epsilon_{ab}
( z^{aA}\partial_A + \partial^{a}_{{\rm expl}} ) 
Q^{b}(z_0)\lpart_B\left. w^B_0 \right]~, \label{s2c2}  \\
S_n[w,\pi]&=&\int dt \left[ \sqrt{g} w^A_3 \pi_A^0
- \pi_A^1~ \tilde{n}_a w^{aA} \right]~,\label{s3c2} 
\eea
where we have introduced an orthogonal unit-vector 
\beq
\tilde{n}_a~:=~\sqrt{g} \epsilon_{ab} n^b~.
\eeq
Together the pair $n^a$ and $\tilde{n}^a$ form an orthonormal basis 
satisfying a completeness relation 
\beq
n^a n_b+ \tilde{n}^a \tilde{n}_b~=~\delta^a_b~.
\eeq
Any quantity $F^a = \theta^a, Q^{a}, z^{aA},w^{aA}, \ldots$, 
carrying $Sp(2)$ indices has projections 
\beq
{}F^{\parallel}~=~n_a F^a~,~~~~~~~~~~~~~~~~~~~~~~~
{}F^{\perp}~=~\tilde{n}_a F^a~,
\label{sp2proj}
\eeq
onto $n^a$ and $\tilde{n}^a$, respectively. They decompose as
\beq
{}F^a~=~n^a F^{\parallel}+ \tilde{n}^a F^{\perp}~.
\eeq
Alternatively, one may think of the above decomposition as a change of 
coordinates $\theta_a \to \theta'_b=\theta_a \Lambda^a{}_b$, where 
\beq
   \theta'_1 ~\equiv~ \theta_{\parallel}~,~~~~~~~~~~~~~~~
\theta'_2 ~\equiv~ \theta_{\perp}~,~~~~~~~~~~~~~~~
\Lambda^a{}_b~=~n^a\delta^1_b + \tilde{n}^a\delta^2_b ~.
\eeq
The unit-vector $\vec{n}$ becomes parallel to the new $1'$-axis, 
and the metric becomes diagonal in the new coordinates:
\beq
  n'_b~=~n_a~\partial^a\theta'_b ~=~ \delta^1_b
~,~~~~~~~~~~~~~~~~~g'_{ad}
~=~\partial^b\theta'_a~ g_{bc}~\partial^c\theta'_d~=~\delta_{ad}~.
\label{ngdelta}
\eeq
If we expand the kinetic part of the action $S$ to the quadratic order,
the $10$ component fields pair off, as indicated in the following 
the diagram 
\beq
\begin{array}{cccc}
z_0&z^{\parallel}&z^{\perp}&z_3 \\
&\updownarrow&\updownarrow&\updownarrow \\
w_3&w^{\perp}&w^{\parallel}&w_0 \\
\updownarrow&\updownarrow \\
\pi^0&\pi^1 
\end{array} \label{diagram2}
\eeq
In other words, the arrows indicate non-zero, off-diagonal entries 
of the kinetic action Hessian.

\subsection{Integration over $\pi^A$}

\noi
It is clear from the third part of the action \e{s3c2} (and in accordance 
with the above diagram), that the integration over the $N\!=\!1$ superfield 
$\pi_A(\theta_{\parallel})=\pi_A^0+\theta_{\parallel}~\pi_A^1$ 
annihilates two components $w^A_3\cong 0$ and $w^{\perp A}\cong 0 $. 
Hence the $N\!=\!2$ superfield
\beq
   w^A~\cong~w^A_0+\theta_{\parallel}~w^{\parallel A}~
\eeq 
reduces to a $N\!=\!1$ superfield. The integrations over the 
superpartners $\pi^0_A$ and  $\pi^1_A$ produce no contribution to the 
measure\footnote{The preservation of measure factors is a very important 
{\em generic} property of a {\em super}field integration, as the component 
fields automatically carry opposite Grassmann statistics. 
This trivial fact will be used so many times in the following, 
that we will not always bother to mention it explicitly.}.
To recapitulate, the field content is now a $N\!=\!2$ superfield $z^A$ and 
a $N\!=\!1$ auxiliary superfield $w^A_0+\theta_{\parallel}~w^{\parallel A}$, 
yielding $2^2+2^1=6$ components for each $A=1,\ldots, 2M$. The partially 
reduced action reads
\bea
   S &\cong &\tilde{S}[z]+\cech{S}[z, w_0, w^{\parallel}]~,  \\
\tilde{S}[z]&=& \int dt \left[ z_0^A \bar{\omega}_{AB}(z_0)~\dot{z}_0^B\right. 
+ \Hf z^{\parallel A} \omega_{AB}(z_0)~ z^{\parallel B}(-1)^{\epsilon_B} 
+ \Hf z^{\perp A} \omega_{AB}(z_0)~ z^{\perp B}(-1)^{\epsilon_B}
\nonumber \\
&&+  \left.  (z^{\parallel A}\partial_A 
+ \partial^{\parallel }_{{\rm expl}}) Q^{\parallel}(z_0)
+ (z^{\perp A}\partial_A + \partial^{\perp }_{{\rm expl}}) 
Q^{\perp}(z_0) \right]~, \\
\cech{S}[z, w_0, w^{\parallel}]
&=&  - \int dt  \left[w^{\parallel A} [  \omega_{AB}(z_0)~z^{\perp B} 
(-1)^{\epsilon_B} + \partial_A Q^{\perp}(z_0) ]\right. 
\nonumber \\
&& + 2 \sqrt{g} w^A_0 ~  \omega_{AB}(z_0)~z^B_3   
+ (-1)^{\epsilon_A} ( z^{\parallel A} z^{\perp B}
-z^{\perp A} z^{\parallel B} )\partial_B \omega_{AC}(z_0)~ w^C_0   
\nonumber \\
&& + [ ( z^{\perp A}\partial_A + \partial^{\perp }_{{\rm expl}} ) 
Q^{\parallel}(z_0) 
- (z^{\parallel A}\partial_A + \partial^{\parallel}_{{\rm expl}} ) 
Q^{\perp}(z_0)] \lpart_B \left. w^B_0 \right]~.
\eea

\subsection{Classical Equations}
\label{seclaseq}

\noi
Now let us vary the action. The $z^A_3$ field only appears at one place 
in the action and to the linear order. Hence, the variation wrt.\ $z^A_3$ 
annihilates $w^A_0\cong 0$:
\beq
\delta z^A_3:~~ w^A_0 ~\cong~0~. 
\eeq
Variation wrt.\ $z^{\parallel A}$ and $z^{\perp A}$ yield their own 
equations,
\bea
\delta z^{\parallel A}:&&
z^{\parallel A}~=~\{ Q^{\parallel}(z_0),z_0^A \}_{PB}+ {\cal O}(w_0)~, \\ 
\delta z^{\perp A}:&&
z^{\perp A}~=~\{ Q^{\perp} (z_0),z_0^A \}_{PB}+w^{\parallel A}+ 
{\cal O}(w_0)~, \label{zperpvar}
\eea
respectively, with the notable appearance of a ``reaction force'' 
$w^{\parallel A}$ in the equation for $z^{\perp A}$. On the other hand 
the $w^{\parallel A}$ field only appears linearly in the action. The variation 
wrt.\ $w^{\parallel A}$ enforces the correct equation for $z^{\perp A}$
\beq
\delta w^{\parallel A}:~~~~
z^{\perp A}~=~\{ Q^{\perp} (z_0),z_0^A \}_{PB}+ {\cal O}(w_0)~.
\label{wparvar} 
\eeq
Note the difference between the $z^{\parallel A}$ and the $z^{\perp A}$ 
sector: The $z^{\parallel A}$ appears Gaussian and free of reaction 
forces, while the $z^{\perp A}$ is constrained. By comparing the two 
\eqs{zperpvar}{wparvar}, we conclude that the reaction force vanishes 
on-shell:
\beq
w^{\parallel A}~=~{\cal O}(w_0)~. 
\eeq
The vanishing of the reaction force is a result of a remarkable balance 
between the tilde part $\tilde{S}$ and the check part $\cech{S}$ of the 
action. {}Finally, variation wrt.\ $w^A_0$ produces an equation for $z^A_3$.
After substitution of appearances of $z^{\parallel A}$, $z^{\perp A}$ 
and $w^A_0$ with their respective equations of motion, 
the $z_3$-equation takes the form:
\beq
z^A_3 ~\cong~ \Hf \epsilon_{ab} \left(
\{Q^a(z_0),\{Q^b(z_0),z^A_0\}_{PB}\}_{PB}
+\{\partial_{{\rm expl}}^a Q^b(z_0),z^A_0\}_{PB}\right)~.
\label{z3transfq}
\eeq 

\subsection{Reduction to $N=0$}
\label{secred20}

\noi
Of the six remaining component fields, let us integrate out $z^A_3$, 
$w^A_0$, $w^{\parallel A}$ and $z^{\perp A}$, leaving $z^A_0$ and 
$z^{\parallel A}$ (See Diagram~\ref{diagram2}). This produces no measure 
factor contributions. The action becomes the $N\!=\!1$ action \e{sc1} 
\beq
S~\cong ~\tilde{S}[z_0,z^{\parallel}] 
+ \int dt \left[  \Hf \{ Q^{\perp}(z_0), Q^{\perp}(z_0) \}_{PB}
+ \partial^{\perp }_{{\rm expl}} Q^{\perp}(z_0)\right]~,
\label{s21shift}
\eeq 
shifted with an additional contribution to the Hamiltonian from the 
perpendicular sector. An integration over the remaining superpartner 
$z^{\parallel A}$ reproduces the $N\!=\!0$ path integral with the 
Hamiltonian \e{ham2}. 

\noi
Note that \e{s21shift} is {\em not} a manifest $N\!=\!2$ to $N\!=\!1$ 
reduction, because of the $N\!=\!0$ shift-term.
Integrating out the anti-BRST symmetry in a $Sp(2)$ theory will encode
{\em half} of the Hamiltonian in a $N\!=\!1$ action and {\em half} 
of the Hamiltonian will be deposited outside in a $N\!=\!0$ term.
This is the inevitable consequence of a BRST and an anti-BRST symmetry, 
which share the same Hamiltonian, or equivalently, that there are
two $\theta$'s but only one $t$. The upshot is that the $N\!=\!2$ theory
can only be compared with the fully reduced $N\!=\!0$ theory, and
not with intermediate stages.

\setcounter{equation}{0}
\section{$N$ Supercharges}
\label{genn}

\noi
We now generalize the construction to arbitrary number $N$ of supercharges
and provide further details, that we previously skipped or glozed over. 
The corresponding $\theta$-space is a $N$-dimensional manifold with 
global, real, Grassmann-odd coordinates $\theta_1, \ldots, \theta_N$.

\subsection{Symplectic $z$-space}
\label{secsympz}

\noi
The superfield $z^{A}$, $A=1,\ldots, 2M$, has Grassmann-parity 
$\epsilon_A$ and takes value in a field-theoretic phase space. 
This phase space is endowed with a closed, non-degenerated, symplectic, 
Grassmann-even $2$-form
\beq
    \omega ~=~ \Hf ~dz^A ~\omega_{AB}~dz^B~,~~~~~~~~~~~~~~~~~~\omega_{AB} 
~=~  (-1)^{(\epsilon_A+1)(\epsilon_B+1)} \omega_{BA}~.
\eeq
The closeness relation $d\omega=0$ reads in components
\beq
  \sum_{{\rm cycl.}~A,B,C} (-1)^{(\epsilon_A+1)\epsilon_C}
\partial_A \omega_{BC}~=~0~,~~~~~~~~~~~~~~~~~~
\partial_A \equiv {\partial \over {\partial z^A}}~. 
\eeq
The superfields can be expanded in $2^N$ component fields 
\beq
z^A(t,\theta) 
~=~ z^A_0(t) + \theta_a ~z^{aA}(t) +  \theta_a \theta_b ~z^{abA}(t)
+ \ldots + (\theta_1 \ldots \theta_N)~z^{1\ldots N,A}(t)~.
\label{z}
\eeq

\subsection{Symplectic $\theta$-space}
\label{symplectheta}

\noi
We now postulate that as a first principle the $\theta$-space should 
be {\em symplectic}, \ie endowed with a closed, non-degenerated, 
symplectic, Grassmann-even $2$-form
\beq
   g~=~\Hf ~d\theta_a~g^{ab}~d\theta_b~,
\eeq
with a reality condition $(g^{ab})^*=g^{ab}$. We will sometimes refer 
to the $\theta$-space as a fermionic world volume.
The closeness relation $dg=0$ reads in components
\beq
  \sum_{{\rm cycl.}~a,b,c} \partial^a g^{bc}~=~0~,~~~~~~~~~~~~~~~~~~
\partial^a \equiv {\partial \over {\partial \theta_a}}~. 
\eeq
The full potential and naturalness of this definition will become 
completely clear after the construction of the superderivative 
in Section~\ref{susyn}.
We begin by discussing some of its immediate consequences.  
{}First of all, the symplectic $2$-tensor turns out to be symmetric,   
\beq
 g^{ab}~=~g^{ba}~,
\eeq 
because of the Grassmann nature of the $\theta$-parameters. 
This remarkable virtue of supermathematics, \ie that a fermionic 
symplectic manifold can superficially ``appear'' Riemannian, is one of 
the main points that we want to state. {}From our experience with the 
$N\!=\!2$ case, we already know that a symmetric metric $g^{ab}$ is 
precisely what is needed in the Susy algebra construction \e{susyalg2}.
The metric 
\beq
g^{ab}(\theta)~=~g^{ab}_0+\theta_c g_1^{c,ab}+
\theta_c\theta_d g_2^{cd,ab}+\ldots
\eeq 
may depend on the $\theta_a$'s but not on the time variable $t$, as this 
would generate unwanted terms in the Susy algebra, see \eq{susyalg} below. 

\noi
Despite the Riemannian ``look'', it is crucial that the $\theta$-space 
metric is of symplectic rather than of Riemannian origin, so that 
powerful symplectic techniques apply:
\begin{enumerate}
\item
{}First of all, a global Poincar\'e Lemma is at our disposal, \ie there 
exists a global, Grassmann-even $1$-form, known as a symplectic potential,
\beq
\Hf \V~=~\Hf \V^a d\theta_a~,~~~~~~~~~~~~~~~~~~~~(\V^a)^*~=~\V^a~.
\label{symppot}
\eeq
such that
\beq
g~=~\Hf d\V~,~~~~~~~~~~~~~~~~~~~~~~
g^{ab}~=~\Hf(\partial^a \V^b+\partial^b \V^a)~.
\label{gv}
\eeq
It is proved in Appendix \ref{appsymp} that for any choice of potential,
there exist fermionic constants $\theta_a^0$ such that
\beq
 \V_a~:=~g_{ab}\V^b~=~
(\theta_a-\theta_a^0) + {\cal O}((\theta\!-\!\theta^0)^2)~.
\eeq
The fermionic constants $\theta_a^0$ are theoretically needed for 
translational invariance in $\theta$-space. They may in practice be set 
to zero. (A reader only interested in the flat case, may identify $\V_a$ 
and $\theta_a$. This is why we choose the unconventional factor $\Hf$ 
in \eqs{symppot}{gv}.)

\item
Secondly, we have a global version of the Darboux theorem, \ie there 
exists global coordinates such that $g^{ab}$ is constant. Moreover, the 
real metric $g^{ab}$ may be chosen to be a diagonal matrix with diagonal 
entries equal to $\pm1$. In fact, the metric has an invariant $(p,q)$ 
signature, $p+q=N$. This is in accordance with a theorem known 
in the mathematical Literature as the Sylvester Law of Inertia. 
For simplicity, we assume in the following that the metric 
$g^{ab}$ is positive definite.

\item
Thirdly, the Liouville theorem, ``Hamiltonian flows are divergenceless'',
will play an important role later on in our construction. The canonical 
world volume measure factor can be written $\sqrt{g}$, where 
$g={\rm det}(g_{ab})$. The Liouville theorem may conveniently be recasted as
\beq
   \Delta~:=~\frac{1}{\sqrt{g}}\partial^a\sqrt{g}g_{ab}\partial^b~=~0~.
\label{deltaiszero}
\eeq
Superficially, this appears as the vanishing of a Laplace-Beltrami operator.
\end{enumerate}

\subsection{Poisson Bracket}

\noi
The inverse symplectic matrix $\omega^{AB}$ is a Poisson bi-vector, which 
gives rise to a Poisson bracket in the exterior algebra ${\cal A}$ of 
world volume forms
\beq
 \{F,G\}_{PB}~=~F\lpart_A ~\omega^{AB}~ \partial_{B}G
~=~-(-1)^{\epsilon(F)\epsilon(G)+p(F)p(G)}\{G,F\}_{PB}
~,~~~~~~~~~~~F,G \in {\cal A}~,
\eeq
where $p$ denotes the world volume form-degree and $\epsilon$ denotes 
the Grassmann-parity. (The form-basis $d\theta_a$ are passive spectators
to the Poisson bracket.) The Jacobi identity reads
\beq
 \sum_{{\rm cycl.}~F,G,H} (-1)^{\epsilon(F)\epsilon(H)+p(F)p(H)}
\{\{F,G\}_{PB},H\}_{PB}~=~0~.
\eeq
Also note that the Poisson bi-vector $\omega^{AB}=\omega^{AB}(z(t,\theta))$ 
does not have explicit $t$ and $\theta$ dependence.

\subsection{Susy Algebra}
\label{susyn}

\noi
We use the symplectic potential $\V^a$ to define the superderivative 
\beq
D^a~=~\partial^a+\norma \V^a \partial_t~,~~~~~~~~~~~~~~~~~~
\partial_t \equiv {\partial \over {\partial t}}~.
\label{superderiv}
\eeq
The exterior superderivative becomes 
\beq
D~:=~d\theta_a ~D^a~=~d-\norma \V \partial_t~,
\label{extsuperderiv}
\eeq
while the Susy algebra itself reads
\beq
D^2=- \twonorma g ~\partial_t~,~~~~~~~~~~~~~~~~~~
  [D^a,D^b]~\equiv~D^{\{a}D^{b\}}~=~ \twonorma g^{ab} \partial_t~,
\label{susyalg}
\eeq
in $2$-form notation and in component notation, respectively.
Note in particular that the exterior superderivative is not nilpotent,
but acts as a square root of time translations.
The metric $g^{ab}$ reappears in the Susy algebra, because of its dual
role as a symplectic field strength (cf.\ \eq{gv}). This is the real reason 
the $\theta$-space is promoted to be a symplectic manifold.

\subsection{Equations of Motion}

\noi 
Superevolution of a quantity $F=F(z(t,\theta);t,\theta)$ 
is governed by $N$ supercharges $Q^a=Q^a(z(t,\theta);t,\theta)$, 
$a=1, \ldots, N$, which can be neatly packed into a Grassmann-even 
supercharge $1$-form
\beq
  Q~:=~d\theta_a ~Q^a~.
\eeq 
At the classical level, the equations of motion for a world volume scalar 
$F$ is
\beq
DF~=~\{Q,F\}_{PB}+ D_{{\rm expl}} F~,
\label{qeom}
\eeq
or in components
\beq
D^a F~=~\{Q^a,F\}_{PB}+ D_{{\rm expl}}^a F~,
\label{qeomc}
\eeq
where the subscript ``expl'' as usual denotes explicit differentiation.

\noi
As an aside, if $F$ in \eq{qeomc} is {\em not} a scalar, but if $F$ 
transforms covariantly as a tensor under world volume reparametrizations, 
one should replace $\partial^a$ with a covariant derivative $\nabla_a$. 
Hence the covariant superderivative reads
\beq
 D^a~=~\nabla^a+\norma \V^a \partial_t~.
\label{covsuperderiv}
\eeq 
This of course depends on the specific choice of connection.
Unfortunately, in contrast to Riemannian manifolds, 
which can always be endowed with a Levi-Civita connection, 
symplectic spaces do not have a canonical choice of connection.
So in other words, in order to give a covariant and consistent recipe for
the equations of motion \e{qeomc} for a non-scalar tensor $F$, we
should assume that the $\theta$-space comes equipped with a connection.
{}For technical reasons, we should assume that the connection 
is torsion free, and other conditions to be discussed 
elsewhere, as it would be out of scope here.

\noi
We now indicate $3$ different derivations of the Susy algebra of the 
$Q^a$'s:
\begin{enumerate}
\item
{}First of all, let us derive the Susy algebra, while only referring to 
manifest scalar objects. It is here that the form notation comes in 
extra handy, as the exterior superderivative $D$ has the nice property, 
that it takes a world volume scalar to a world volume scalar. Hence 
we may apply \e{qeom} repeatedly on itself, without having to rely on 
a specific connection. (Note in particular that this is not so for the 
superderivative $D^a F$, which is {\em not} a world volume scalar.) 
Applying the equations of motion \e{qeom} successively, and using the 
Jacobi identity, we get
\beq
-  \twonorma g~\partial_t F~=~ D^2 F~=~
\left\{\Hf \{Q,Q\}_{PB}+  D_{{\rm expl}} Q ,F\right\}_{PB}
-  \twonorma g ~\partial^{{\rm expl}}_t F~.
\eeq
Written out in components
\beq 
 \twonorma g^{ab} \partial_t F~=~[D^a,D^b]F~=~
\left\{\{Q^a,Q^b\}_{PB}+ D_{{\rm expl}}^{\{a} Q^{b\}} , F\right\}_{PB}
+  \twonorma g^{ab}\partial^{{\rm expl}}_t F~.
\label{doublecom}
\eeq
\item
Secondly, if one do not like forms, to justify the component expression 
\e{doublecom} directly in components, one merely has to check that 
\e{doublecom} transforms covariantly under change of coordinates. 
\item
Thirdly, one can derive \eq{doublecom} in components by applying the 
superderivative $D^a$ twice and {\em then} symmetrize, with the implicit 
understanding that the $D^a$ in \eq{doublecom} stands for the 
{\em covariant} superderivative \e{covsuperderiv}. However, if the 
connection is torsion free, it is easy to see that the Christoffel 
symbols drops out of the symmetrized combination in \eq{doublecom}, 
so one may replace $D^a$ with ordinary superderivatives 
\e{superderiv}, in agreement with the two previous derivations.
\end{enumerate}

\subsection{Hamiltonian}

\noi
Contracting on both sides of \eq{doublecom} with the inverse metric
$g_{ab}$, we get the equation for time evolution
\beq
\partial_t F~=~-\{H,F\}_{PB}+\partial^{{\rm expl}}_t  F~,
\label{teom}
\eeq
where we define the Hamiltonian as
\beq
-H~:=~\Hf g_{ab} 
\left( \{Q^a,Q^b\}_{PB}+ D_{{\rm expl}}^{\{a}Q^{b\}} \right)~.
\label{hamn}
\eeq
Note that $H$ depends explicitly on the metric $g_{ab}$.

\subsection{Integrability}

\noi
Similarly, the $\theta$-evolution is governed by the following 
equations of motion 
\beq
\partial^a F~=~\{\Omega^a,F\}_{PB}+\partial_{{\rm expl}}^a  F
~,~~~~~~~~~~~~~~~~~~
dF~=~\{\Omega,F\}_{PB}+d_{{\rm expl}} F~,
\label{thetaeom}
\eeq
with generators
\beq
\Omega^a~:=~Q^a+\norma \V^a H~,~~~~~~~~~~~~~~~~~~~~
\Omega~:=~d \theta_a~\Omega^a~=~Q-\norma \V H~.
\eeq
The integrability conditions $[\partial_t,d]F=0$ and $d^2F=0$ for 
\eqs{teom}{thetaeom} are zero-curvature equations
\bea
\{\Omega,H\}_{PB} + d_{{\rm expl}} H
+ \partial^{{\rm expl}}_t \Omega   &=& 0 \label{zerocurv1f} \\
\Hf \{\Omega,\Omega\}_{PB} 
+ d_{{\rm expl}} \Omega  &=& 0~.
\label{zerocurv2f}
\eea

\subsection{Two Sectors: Tilde and Check}
\label{twosec}

\noi
In any dimension $N$ there are two $1$-forms readily at our disposition. 
They are the symplectic potential $\V^a$ and its Hodge-like dual 
\beq
\tilde{\V}^a~:=~D^a  \left( \frac{\V^N}{\sqrt{g}} \right)
 ~=~\partial^a \delta^N(\theta\!-\!\theta^0)~,
\eeq
respectively. (See the Appendices for further details.) As usual, the 
metric $g_{ab}$ raises and lowers indices. Hence we may construct two 
types of scalar supercharges
\bea
\tilde{Q}&:=&\tilde{\V}_a Q^a 
~=~\tilde{\V}_a \Omega^a + (-1)^N \delta^N(\theta\!-\!\theta^0)~H 
\label{tildeq} \\
\cech{Q}&:=&\V_a Q^a~=~\V_a \Omega^a~.
\label{cechq}
\eea
and two types of superderivatives\footnote{$\cech{D}$ and $\cech{Q}$ 
were denoted $D$ and $Q$, respectively, in \Ref{BD}.}
\bea
\tilde{D}&:=&\tilde{\V}_a D^a 
~=~\tilde{\V}_a \partial^a 
-(-1)^N \delta^N(\theta\!-\!\theta^0)~\partial_t 
 \label{tilded} \\
\cech{D}&:=&\V_a D^a~=~\V_a \partial^a~.
\label{cechd}
\eea
In the last equality in both \eqs{cechq}{cechd} we used the fact 
that $\V^a g_{ab} \V^b=0$.
The equations of motion \eq{qeom} for $D^a$ imply the corresponding
equation of motion for $\tilde{D}$ and $\cech{D}$:
\bea
\tilde{D}F&=&\{\tilde{Q},F\}_{PB}+ \tilde{D}_{{\rm expl}} F 
\label{tildeeom} \\
\cech{D}F&=&\{\cech{Q},F\}_{PB}+ \cech{D}_{{\rm expl}} F~.
\label{cecheom}
\eea
We show in the next two Sections~\ref{tildesec}-\ref{cechsec}, that the 
opposite is also true, \ie that the two \eqs{tildeeom}{cecheom} imply the 
\eq{qeom}.

\subsection{Tilde Sector}
\label{tildesec}

\noi
{}First multiply the tilde equation \e{tildeeom} with $\V^a$ 
(or the check equation \e{cecheom} with $\tilde{\V}^a)$:
\beq
  \delta^N(\theta\!-\!\theta^0) ~D^a F
~=~\delta^N(\theta\!-\!\theta^0)~\{Q^a,F\}_{PB}
+ \delta^N(\theta\!-\!\theta^0)~D_{{\rm expl}}^a F~,
\eeq
leading to
\beq
d F(\theta^0)~=~\{\Omega(\theta^0),F(\theta^0)\}_{PB}
+d_{{\rm expl}} F(\theta^0)~.
\label{thetaeom0}
\eeq
Our aim is now to derive a differentiated version
\beq
0~=~d^2F ~=~ d\{\Omega(\theta^0),F(\theta^0)\}_{PB}
+d d_{{\rm expl}} F(\theta^0)~.
\label{thetaeom0d}
\eeq
One may not proceed by direct differentiation, because \eq{thetaeom} is 
only known for $\theta_a=\theta^0_a$. Instead substitute $F$ with 
$\{\Omega,F\}_{PB}+d_{{\rm expl}} F$ in the above \eq{thetaeom0}:
\bea
d\{\Omega(\theta^0),F(\theta^0)\}_{PB}+dd_{{\rm expl}} F(\theta^0)&=&
\{\Omega(\theta^0),\{\Omega(\theta^0),F(\theta^0)\}_{PB}  \}_{PB} \cr
&&+\{\Omega(\theta^0), d_{{\rm expl}} F(\theta^0) \}_{PB}
+d_{{\rm expl}}\{\Omega(\theta^0),F(\theta^0)\}_{PB} \cr 
&=&0~,
\eea
where the Jacobi identity and a integrability condition \e{zerocurv2f} 
were applied in the last equality.

\noi
Next Berezin integrate the lhs.\ of the tilde equation \e{tildeeom}.
After integrating by part and using the Liouville theorem \e{deltaiszero}, 
we get 
\beq
\int  \sqrt{g} ~ d^N\theta~ \tilde{D}F~=~
-(-1)^N ~\partial_t F(\theta^0)~.
\eeq
Similarly, after use of \eqs{thetaeom0}{thetaeom0d}, 
the rhs.\ of the tilde equation \e{tildeeom} becomes
\beq
\int  \sqrt{g} ~ d^N\theta 
\left( \{\tilde{Q},F\}+\tilde{D}_{{\rm expl}} F\right)~=~
(-1)^N ~\{H(\theta^0) ,F(\theta^0)\}
-(-1)^N ~\partial^{{\rm expl}}_t F(\theta^0) ~.
\eeq
Hence we have derive the equations of motion \es{thetaeom}{teom} 
(and thereby the superequation \e{qeom}) for $\theta_a=\theta^0_a$.

\subsection{Check Sector}
\label{cechsec}

\noi
Imagine that we are given initial conditions 
$z^A(t_0,\theta_0)=z^A_{00}$. It is shown above that one 
can determine uniquely $z^A(t,\theta_0)$ for arbitrary times 
$t$ along the line $\theta_a=\theta_a^0$. We would like to extend
the solution to arbitrary $\theta_a\neq\theta_a^0$, at some arbitrary
but fixed time $t$.
The key to solve the check equation $\cech{D}z^A=\{\cech{Q},z^A\}_{PB}$,  
is to keep track of powers of $\theta_a-\theta_a^0$. 
Let $F_{[n]}$ be the part of a quantity $F$ that contains $n$ powers 
of $\theta_a-\theta_a^0$. 
We expand the relevant quantities accordingly:
\bea
z^A&=&\sum_{n=0} z^A_{[n]}~,~~~~~~~~~~~~~~~~~~
z^A_{[0]}~:=~z^A(t,\theta_0)~, \\
\{\cech{Q},z^A\}_{PB}&=&\sum_{n=1} \{\cech{Q},z^A\}_{PB}^{[n]}~,\\
\cech{D}&=&\sum_{n=0} \cech{D}_{[n]} ~.
\eea
All sums truncate because the $\theta$'s are Grassmann-odd.
The leading contribution to $\cech{D}$ is the conformal operator
\beq
\cech{D}_{[0]}~=~(\theta_a-\theta_a^0) \partial^a~.
\eeq
The operator $\cech{D}_{[0]}$ preserves and counts the powers of 
$\theta_a-\theta_a^0$. Applying this on the check equation, we get 
\beq
  n z^A_{[n]}~=~ \cech{D}_{[0]}z^A_{[n]}
~=~\{\cech{Q},z^A\}_{PB}^{[n]}-\sum_{k=1}^{n}\cech{D}_{[k]} z^A_{[n-k]}~.
\label{triz}
\eeq
It is easy to solve for $z^A_{[n]}$. 
We can do it successively, for increasing $n>0$,
because \eq{triz} is of a triangular form. Recall namely that $\cech{Q}$ 
contains explicitly one power of $\theta_a-\theta_a^0$. 
Therefore, for a given $n>0$, the rhs.\ can only depend on previous components
$z^A_{[0]}, \ldots, z^A_{[n-1]}$. 

\noi
This shows that there is enough information in the tilde and check sector
\eqs{tildeeom}{cecheom} combined to construct a unique solution 
$z^A(t,\theta)$ for arbitrary $t$ and $\theta$. Will this solution $z$ 
satisfy the $D^a$ equations of motion \eq{qeom} as well? Yes, because 
a solution $z'$ to \eq{qeom} (with the same initial condition) is trivially 
also a solution to \eqs{tildeeom}{cecheom}. By uniqueness $z=z'$.

\subsection{Action}

\noi 
Our action is a natural generalization of the $N\!=\!2$ action \e{ss2},
\bea
  S&=&\tilde{S}[z]+\cech{S}[z,w]+S_n[w,\pi]~,  \label{ssn} \\
\tilde{S}[z]&=&-(-1)^N \int dt \sqrt{g} ~d^N\theta
\left[ z^A \bar{\omega}_{AB}(z)~\tilde{D}z^B(-1)^{N \epsilon_B} 
+\tilde{Q}(z)\right]~, \label{s1n} \\
\cech{S}[z,w]&=&\int dt \sqrt{g} ~d^N\theta~w^A 
\left[  \omega_{AB}(z)~\cech{D}z^B + \partial_A\cech{Q}(z)\right]~, \\
S_n[w,\pi]&=&\int dt \sqrt{g} ~d^N\theta~w^A(\theta)~
\pi_A(n^a \theta_a)~.
\eea
Here $z^A$ and $w^A$ are superfields, while $\pi_A$ is a $N\!=\!1$ 
auxiliary superfield. Of these, $z^A$ and $\pi_A$ carry the same 
Grassmann-parity $\epsilon_A$, while $w^A$ carries Grassmann-parity 
$\epsilon_A\!+\!N$. The real unit-vector $n^a=n^a(\theta)$ represents an 
explicit choice of direction in the $N$-dimensional $\theta$-space. 
We will show how the unit-vector $n^a$ points out a single superpartner 
$z^{\parallel A}:=n_a z^{aA}$, which appears to the Gaussian order 
in the action, while all the remaining $N\!-\!1$ orthogonal $z^{aA}$ 
superpartners are constrained. The $\bar{\omega}_{AB}$ is defined as
\beq
\bar{\omega}_{AB}:=~\left(z^C \partial_C+2\right)^{-1}\omega_{AB}
~=~\int_0^1 \alpha~d\alpha~\omega_{AB} (\alpha z)~.
\label{omegabar}
\eeq 
We claim that the variation of the action yields the 
equations of motion
\bea
  \tilde{D}z^A&=&\{\tilde{Q},z^A\}_{PB}~, \label{tildeneom} \\
   \cech{D}z^A&=&\{\cech{Q},z^A\}_{PB}~,  \label{cechneom} \\
      w^A &=& 0~,  \label{wneom} \\
 \pi_A &=& 0   \label{pineom}~.
\eea
We prove parts of this below, in particular that the full superfield 
$w^A$ vanishes on-shell: $w^A \cong 0$. 
As a corollary to this, we immediately deduce that the tilde part 
$\tilde{S}$ alone is responsible for the equation of motion for 
the superfield $z^A$. On the other hand, it is straightforward to perform a 
manifest superfield variation of $\tilde{S}$. This is to a large extend
insensitive to the number of superpartners, 
and the resulting equation is \eq{tildeneom}.

\noi
It is proved in Appendix~\ref{appsymp} that the (lowered) symplectic 
potential $\V_a$ has a unique fermionic zero $\theta^0_a$.
We assume from now on that $\theta^0_a$ is shifted to $\theta^0_a=0$, 
either by redefining the $\V_a$ or the $\theta_a$. 

\noi
The first part of the action $\tilde{S}$ turns out to consist of terms 
proportional to either the delta function $\delta^N(\theta)$ itself 
or its first derivatives. Now integrate the latter type of terms by part.
After use of the Liouville Theorem $\partial^a(\sqrt{g}g_{ab})=0$, 
we arrive at a component expression
\bea
\tilde{S}[z]&=&\int dt \left[ z_0^A \bar{\omega}_{AB}(z_0)~\dot{z}_0^B\right.
+ \Hf  z^{aA}g^0_{ab} \omega_{AB}(z_0)~ z^{bB}(-1)^{\epsilon_B}
\nonumber \\
&& +\left. g^0_{ab}(z^{aA}\partial_A+\partial^a_{{\rm expl}})Q^b(z_0)\right]~.
\label{s1cn} 
\eea
It is hard to provide useful component expressions for the second part 
$\cech{S}$. Instead, we will see below how an inductive approach may yield
manageable expressions (cf.\ Section~\ref{reducsec}). 
The third part reads in components
\beq
S_n[w,\pi]~=~\int dt  \sqrt{g_0} \left[ w^A_{2^N-1} \pi_A^0
-(-1)^{[\frac{N+1}{2}]} \pi_A^1~ n^{a_1}_0\epsilon_{a_1,\ldots,a_N}~
w^{a_2,\ldots,a_n,A} \right]~,
\label{s3cn} 
\eeq
if $g=g_0$ is constant. In the non-constant case, there will be 
additional terms proportional to derivatives of the metric.

\noi
We should mention that it is always possible to go to new 
$\theta'_a$-coordinates, such that the unit-vector $\vec{n}$ becomes 
parallel to the new $1'$-axis, and the metric becomes the unit matrix 
(cf.\ \eq{ngdelta}). However, here we continue to work with as general 
coordinates as possible.

\subsection{Case $N=1$}

\noi
In the $N\!=\!1$ case, there are only two discrete choices for the 
unit-vector $n^1=\pm \sqrt{g^{11}}$, corresponding to the two points 
on a zero-sphere $S^0$. The second and third part of the action read 
in components
\bea
\cech{S}[z,w]&=&-\int dt \sqrt{g_0} ~w^A_0\left[ 
\omega_{AB}(z_0)~z^{1B}(-1)^{\epsilon_B}+\partial_A Q^1(z_0)\right]~,
\label{scech1} \\
S_n[w,\pi]&=&\int dt \left[ \sqrt{g_0}~  w^A_1 \pi_A^0
\pm \pi_A^1 w^A_0 \right]~,
\label{s3cn1}  
\eea
respectively. A closeness relation $\partial^1g^{11}=0$ was used to
derive the component expression for the third part $S_n$.
An $N\!=\!1$ superfield integration over $\pi_a$ annihilates 
$w^A$ completely, so that the action reduces to merely the $z$-sector
\beq
S ~\cong~\tilde{S}[z]~.
\eeq
If we go to flat coordinates $g^{11}=1$, we arrive at the usual 
$N\!=\!1$ action \e{s1} with a Gaussian superpartner $z^{1A}$.

\subsection{Reduction from $N$ to $N-1$}
\label{reducsec}

\noi
Since we have already covered the $N\!=\!1$ case we may assume $N > 1$. 
The reductive step consists of integrating out a $\theta$-direction in 
the $N$-dimensional $\theta$-space. The direction may be chosen arbitrarily, 
as long as it is orthogonal to the given $n^a$ direction. By performing a 
change of coordinates $\theta_a \to \theta'_b$, if necessary, one may without 
loss of generality assume that the direction, that is integrated out, 
is $\theta_1$, and that $n^a$ and $g^{ab}$ are of the block form
\beq
n^a~=~\left[  \begin{array}{c}
0 \cr \hline \cr
n'^a \cr \cr
\end{array} \right]~,~~~~~~~~~~~~~~~~~~~~~
g^{ab}~=~\left[  \begin{array}{c|ccc}
1&0&\cdots&0 \cr \hline
0 \cr
\vdots&&g'^{ab} \cr
0 \cr
\end{array} \right]~.
\eeq 
A prime denotes quantities associated with the $N\!-\!1$ remaining 
directions. For instance, $\theta':=(\theta_2,\theta_3,\ldots,\theta_N)$.
One may take the symplectic potential and its dual to be of the form
\beq
\V^a~=~\left[  \begin{array}{c}
\theta_1 \cr \hline \cr
\V'^a \cr \cr
\end{array} \right]~,~~~~~~~~~~~~~~~~~~~~~
\tilde{\V}^a~=~\partial^a\delta^N(\theta)
~=~\left[  \begin{array}{c}
\delta^{N-1}(\theta')  \cr \hline \cr
-\theta_1 \tilde{\V}'^a \cr \cr
\end{array} \right]~,
\eeq 
respectively, so that the tilde and check superderivatives and 
supercharges decompose as
\bea
\cech{D}&=&\V_a \partial^a~=~\cech{D}'+\theta_1 \partial^1~, 
\label{cechdreduc} \\ 
\tilde{D}&=&\tilde{\V}_a \partial^a 
-(-1)^N \delta^N(\theta)~\partial_t 
~=~\delta^{N-1}(\theta')~\partial^1 -\theta_1\tilde{D}'~,
\label{tildedreduc} \\
\cech{Q}&=&\V_a Q^a~=~\cech{Q}'+\theta_1 Q^1~,
\label{cechqreduc} \\
\tilde{Q}&=&\tilde{\V}_a Q^a 
~=~\delta^{N-1}(\theta')~Q^1 -\theta_1\tilde{Q}'~, 
\label{tildeqreduc} 
\eea
respectively.
The superfields themselves decompose as
\beq
\begin{array}{rclcl}
z^A(\theta)&=&z'^A(\theta')+\theta_1 \hat{z}^{A}(\theta')
&~~\longrightarrow~~&z'^A(\theta')~, \\
w^A(\theta)&=&\hat{w}^A(\theta')+\theta_1 w'^A(\theta')
&~~\longrightarrow~~&w'^A(\theta')~, \\
\pi_A(n^a \theta_a)&=&\pi_A(n'^a \theta'_a)~.
\end{array}
\eeq
In detail, to go from $N$ to $N\!-\!1$ number of $\theta$'s
in the path integral, the hat superfields $\hat{z}^A(\theta')$ and 
$\hat{w}^A(\theta')$ should be integrated out, while the superfields 
$z'^A$, $w'^A$ and $\pi_A$ should be kept.  
Note that the Grassmann parities of the $w^A$ and $w'^A$ fields are opposite.
We do not touch the $\pi_A$ field, as $n^1=0$. (Recall that the 
$\theta$-direction, that is integrated out, is orthogonal to $n^a$.) 
Therefore, the third part $S_n$ is unaltered by the reduction
\beq
 S_n[w,\pi]-S'_n[w',\pi]~=~0~.
\eeq
The change in the first action part is
\beq
\tilde{S}[z]-\tilde{S}'[z']~=~
\int dt\left[\Hf z^{1A} \omega_{AB}(z_0)~ z^{1B}(-1)^{\epsilon_B}
+(z^{1A} \partial_A + \partial^1_{{\rm expl}}) Q^1(z_0)\right]~.
\label{deltatilde} 
\eeq
This follows either from the component expression \e{s1cn},
or one may give a manifest $N\!-\!1$ superfield derivation by applying 
the decomposition formulas \es{tildedreduc}{tildeqreduc} in the
superfield action \e{s1n}.
The change in the second action part becomes
\bea
\cech{S}[z,w]-\cech{S}'[z',w']&=&
(-1)^N\int dt \sqrt{g'} ~d^{N-1}\theta'~\hat{w}^A\left[ 
\omega_{AB}(z')~(\cech{D}'+1)
\hat{z}^{B}(-1)^{\epsilon_B}\right. \nonumber \\
&&\left.+(-1)^{\epsilon_A}\hat{z}^{C}\partial_C\omega_{AB}(z')~
\cech{D}'z'^B 
 +\partial_A [(\hat{z}^{B} \partial_B + \partial^1_{{\rm expl}}) 
\cech{Q}'(z')+ Q^1(z')]\right]~.
\label{deltacech} 
\eea
Now expand everything in powers $n$  of $\theta'$,
\beq
z'^A~=~\sum_{n=0} z'^A_{[n]}~,~~~~
\hat{z}^A~=~\sum_{n=0} \hat{z}^A_{[n]}~,~~~~
\hat{w}^A~=~\sum_{n=0} \hat{w}^A_{[n]}~,~~~~
\cech{D}'~=~\sum_{n=0}\cech{D}'_{[n]}~,~~~~ \ldots
\eeq
We now make the following

\noi
\underline{Claim:}
{}For all $r=0,\ldots,N\!-\!1$, the variation of $S-S'$ wrt.\ 
$\hat{z}_{[N-1-r]}$ leads to the equations of motion 
$\hat{w}_{[r]}\cong 0$.

\noi
\underline{Induction proof in $r$:}
We argue {\em successively}, for {\em increasing} $r=0,\ldots,N\!-\!1$. 
According to the induction assumption we may discard terms proportional 
to previous components $\hat{w}^A_{[0]},\ldots,\hat{w}^A_{[r-1]}$. 
Moreover, we only have to keep terms that contain $\hat{z}^B_{[N-1-r]}$ 
explicitly. Coincidentally, this leave no room for $\theta'$ appearances 
that are not already accounted for inside either $\hat{w}_{[r]}$ or inside 
$\hat{z}^B_{[N-1-r]}$.

\noi
\underline{Case $r\! <\! N\!-\!1$:}
The relevant parts boil down to only one term 
\beq
\cech{S}[z,w]-\cech{S}'[z',w']~\sim~
(-1)^N\int dt \sqrt{g_0'} ~d^{N-1}\theta'~\hat{w}^A_{[r]}
\omega_{AB}(z_0)~(N-r)
\hat{z}^B_{[N-1-r]}(-1)^{\epsilon_B}~.
\label{deltacechrel} 
\eeq
Variation wrt.\ $\hat{z}^B_{[N-1-r]}$ yields the claim
$\hat{w}^A_{[r]} \cong 0$.

\noi
\underline{The last step $r \!=\! N\!-\!1$:}
We would like to vary wrt.\ $\hat{z}^A_{[0]} \equiv z^{1A}$.
Up to now we have showed that the full $N\!-\!1$ superfield $\hat{w}^A$ 
reduces on-shell to a single top-component,
\beq
 \hat{w}^A~\cong~\theta_2\theta_3 \ldots \theta_N w^{23 \ldots N,A}~.
\eeq
Inserted into the second part of the action, we get
\beq
\cech{S}[z,w]-\cech{S}'[z',w']~\cong~
(-1)^N\int dt \sqrt{g_0'} ~w^{23 \ldots N,A}\left[
\omega_{AB}(z_0)~
z^{1B}(-1)^{\epsilon_B}+\partial_A Q^1(z_0) \right]~.
\eeq
We should not forget contributions from the tilde sector \e{deltatilde}, 
which also depend on $\hat{z}^A_{[0]} \equiv z^{1A}$. Variation 
wrt.\ $w^{23 \ldots N,A}$ enforces the correct equations of motion
\beq
 \delta w^{23 \ldots N,A}:~~z^{1B}~\cong~\{Q^1(z_0),z^B_0\}_{PB}~,
\eeq
while the variation wrt.\ $z^{1B}$ yields the same equation 
\beq
 \delta z^{1B}:~~
z^{1A}~\cong~\{Q^1(z_0),z^A_0\}_{PB}
-(-1)^N \sqrt{g_0'} w^{23 \ldots N,A}~,
\eeq
with an additional reaction force term. We conclude that the top-component
\beq
 w^{23 \ldots N,A}~\cong~0
\eeq
of the superfield $\hat{w}^A$ must vanishes on-shell as well.
Completing the square yields the following shift in the action
\beq
S-S'~\cong~ \int dt \left[  \Hf \{ Q^1(z_0), Q^1(z_0) \}_{PB}
+ \partial^1_{{\rm expl}} Q^1(z_0)\right]~.\label{shiftn}
\eeq
It is now clear how the complete reduction down to $N\!=\!0$ proceeds. 
The outcome is the usual $N\!=\!0$ partition function  
\beq
  {\cal Z}~=~ \int [dz] \exp\left[\Ih S \right]
  ~=~\int [dz_0] ~{\rm Pf}(\omega_{AB}(z_0))~  \exp\left[\Ih \int dt\left( 
z_0^A \bar{\omega}_{AB}(z_0)~\dot{z}_0^B-H(z_0)\right)\right]~.
\label{pathintn}
\eeq
{}From each step of the reduction, there is a classical action shift 
\e{shiftn}. The shifts accumulate in the Hamiltonian of the form \e{hamn}. 
{}Finally, in the very last reduction step from $N\!=\!1$ to $N\!=\!0$,
all references to the choice of unit-vector $n^a$ disappear, 
and the correct Pfaffian measure factor is reinstated.

\subsection{An Example: $N=3$}

\noi
The fact that the new action principle \e{ssn} is formulated for 
arbitrary $N$, makes it is applicable to numerous situations. 
{}For instance, from the perspective of Hamiltonian $Sp(2)$ theories, 
it opens up a tantalizing possibility to have a {\em manifest} 
$2$-dimensional $\theta$-plane inside a $N\!=\!3$ theory, consisting 
of three supercharges $Q^a=\Omega^a+\frac{1}{3}\V^a H$, $a=1,2,3$,
with an explicitly broken 3rd direction $n^a=\delta^a_3$, 
and with a corresponding dummy charge $\Omega^3=0$.
Let us assume the coordinates are flat $\V^a=\theta_a$ for simplicity.
{}From our analysis, we know that the construction will implement 
all the correct equations of motion. If we use the $3$-dimensional 
Levi-Civita symbol to raise and lower indices on superpartners, 
\ie $z_a^A:=\Hf \epsilon_{abc}z^{bcA}$, we may indicate how the 
$2^3+2^3+2^1=18$ component fields pair off inside the kinetic part 
of the action to the quadratic order: 
\beq
\begin{array}{cccccccc}
z_0&z^1&z^2&z^3&z_3&z_2&z_1&z_7 \\
&\updownarrow&\updownarrow&\updownarrow&\updownarrow&
\updownarrow&\updownarrow&\updownarrow \\
w_7&w_1&w_2&w_3&w^3&w^2&w^1&w_0 \\
\updownarrow&&&\updownarrow \\
\pi^0&&&\pi^1 
\end{array}
\eeq
The $N\!=\!2$ superpartners $z^1$, $z^2$ and $z_3$ will be constrained 
to their equations of motion, respectively, while the Gaussian 
integration over $z^3$ will produce the Pfaffian measure factor.

\section{Discussions}

\noi
To recapitulate, we have recasted the equations of motion \e{qeom} 
for superevolution into an equivalent set of two equations, 
a tilde \eq{tildeeom} and a check \eq{cecheom}. The tilde
and check equation of motion lead directly to a corresponding action
and path integral prescription. The third and last ingredient 
is a gauge-fixing term $S_n$, which explicitly single out a direction 
$n^a$ in the $\theta$-space without affecting the physical sector.
One may speculate that the above rearrangement is necessary to disentangle 
the $t$ and the $\theta$ derivatives, and thereby unravel the $z_0$ sector
from its many superpartners. The caveat is of course that
all manipulations should be executed in a supersymmetric manner.

\noi
We saw that a comparison of a supersymmetric theory to some of its 
{\em partially} reduced cousins is tricky, because non-supersymmetric
shift-terms are generated in the action (cf.\ Section~\ref{secred20}). 
However, the reduction scheme is very successful in making contact 
to the {\em fully} reduced $N\!=\!0$ theory.
One may imagine that the ``democracy'' among the $N$ $\theta$-directions 
in the Susy algebra, in which they share one and the same $t$ coordinate, 
makes it hard to partially integrate out $\theta$-directions without  
affecting the common $t$ parameter. 

\noi
In conclusion, we have formulated a simple action principle for a 
superfield formulation of Hamiltonian field theories with $N$ supercharges. 
This action gives rise to a viable path integral of superfields, which 
may easily be reduced to the original sector.

\noi
{\sc Acknowledgment:}~This work has been partially supported by
INTAS grant 00-00262. Also, the work of I.A.B.\ has been partially 
supported by President grant LSS-1578-2003.2 and RFBR grants 
02-01-00930 and 02-02-16944.
The work of K.B.\ has been supported by DOE grant DOE-ER-40173.
I.A.B.\ and K.B.\ would like to thank Poul Henrik Damgaard for 
numerous and fruitful discussions, and the Niels Bohr Institute 
for the warm hospitality extended to them there.  

\vspace{1cm}
\appendix

\setcounter{equation}{0}
\section{Superconventions}

The Berezin integration is defined as
\beq
\int d^N\theta~:=~\partial^N \ldots \partial^1
~=~\frac{1}{N!}\epsilon_{a_1,\ldots,a_N}~\partial^{a_1}\ldots \partial^{a_N}~,
\eeq
where we have the following convention for the Levi-Civita symbol
\beq
\epsilon_{N,\ldots,1}~:=~+1~,~~~~~~~~~~~~~~~~~~~~
\epsilon_{1,\ldots,N}~=~(-1)^{[\frac{N}{2}]}~.
\eeq
An invariant integration measure is
\beq
  \sqrt{g} ~ d^N\theta~,~~~~~~~~~~~~~~~~~
g~:=~{\rm Ber}(g^{ab})~=~{\rm det}(g_{ab})~.
\eeq
The invariant $N$-dimensional Dirac delta function is correspondingly
\beq
 \delta^N(\theta)~:=~ \frac{\theta^N}{\sqrt{g}}~=~\frac{\theta^N}{\sqrt{g_0}}~,
\eeq
where
\beq
\theta^N~:=~\theta_1 \ldots \theta_N
~=~\frac{\epsilon^{a_1,\ldots,a_N}}{N!}~\theta_{a_1}\ldots \theta_{a_N}
~,~~~~~~~~~~~~~~~~~\epsilon^{1,\ldots,N}~:=~+1~.
\eeq 
We may define a $1$-form 
\beq
\tilde{\theta}^{a_1}~:=~\partial^{a_1} \delta^N(\theta)
~=~\frac{\epsilon^{a_1,\ldots,a_N}}{\sqrt{g_0} (N\!-\!1)!}~
\theta_{a_2}\theta_{a_3}\ldots \theta_{a_N}~.
\eeq
{}Fierz' relations:
\bea
 \theta_{a_1}\ldots \theta_{a_N}
&=&(-1)^{[\frac{N}{2}]}\epsilon_{a_1,\ldots,a_N}~ \theta^N~,    \cr\cr
 \theta_a\tilde{\theta}^{b}&=&\delta_a^b~\delta^N(\theta)~.
\eea

\setcounter{equation}{0}
\section{The Symplectic Potential}
\label{appsymp}

\noi
In general, $\V^a$ has a $\theta$-expansion
\beq
\V^a(\theta)~=~\V^a_0+g^{ab}_0 \theta_b + {\cal O}(\theta^2)~,
\label{vexp1}
\eeq 
where $\V^a_0$ are fermionic constants. It is easy to see that the 
equation $\V^a(\theta)=0$ has precisely one solution, which we will denote 
\beq
\theta^0_a~=~-g_{ab}^0 \V^b_0 + {\cal O}(\V_0^2)~.
\eeq
({}For instance, rewrite \eq{vexp1} as a fixed point equation 
$\theta_a=-g_{ab}^0 \V^b_0 + {\cal O}(\theta^2)$, and eliminate all 
$\theta$-appearances recursively on the rhs. This process terminates 
after finite many steps because of the Grassmann nature.) 
Hence we may reorganize $\V^a$ as a polynomial in $(\theta_a-\theta_a^0)$
with no constant term:
\beq
\V^a(\theta)
~=~g^{ab}_0(\theta_b-\theta_b^0) + {\cal O}((\theta\!-\!\theta^0)^2)~,
\label{vexp2}
\eeq 
Therefore, the symplectic field strength 
$g^{ab}(\theta)=g^{ab}_0+{\cal O}(\theta\!-\!\theta^0) $ is a function of 
$(\theta_a-\theta_a^0)$ as well. We conclude in particular that
\beq
g^{ab}(\theta^0)~=~ g^{ab}_0~\equiv~g^{ab}(0)~,
\eeq
which can be traced to the fact that $g^{ab}$ does not depend on the 
fermionic constants $\V^a_0$. Moreover,
\beq
 \V_a~:=~g_{ab}\V^b
~=~(\theta_a-\theta_a^0) + {\cal O}((\theta\!-\!\theta^0)^2)~.
\eeq
A $N$-fold product of the $\V_a$'s leads to a shifted delta function
\beq
 \frac{\V^N}{\sqrt{g}}~=~\frac{\V^N}{\sqrt{g_0}}
~=~ \delta^N(\theta\!-\!\theta^0)~,~~~~~~~~~~~~~~~~~~~~~
\V^N~:=\V_1 \ldots \V_N~.
\eeq
We may define a $1$-form 
\beq
\tilde{\V}^{a_1}~:=~\partial^{a_1} \left( \frac{\V^N}{\sqrt{g}} \right)
~=~\frac{\epsilon^{a_1,\ldots,a_N}}{\sqrt{g_0} (N\!-\!1)!}~
(\theta_{a_2}-\theta_{a_2}^0)(\theta_{a_3}-\theta_{a_3}^0)
\ldots (\theta_{a_N}-\theta_{a_N}^0)~.
\eeq
{}Fierz' relation:
\beq
 \V_a \tilde{\V}^{b}~=~\delta_a^b~\delta^N(\theta\!-\!\theta^0)~.
\label{fierz}
\eeq
One may also give explicit formulas for $\V^a$ in terms of $g^{ab}$ 
by using homotopy operators. A convenient choice of potential, 
satisfying additionally $\V^a\theta_a=0$ and $\V^a|_{\theta=0}=0$, is
\beq
\Hf \V^b~=~\theta_a~\bar{g}^{ab}~,~~~~~~~~~~~~~~~~~
\bar{g}^{ab}~:=~(\theta_a \partial^a+2)^{-1}g^{ab}
~=~\int_0^1 \alpha~d\alpha~ g^{ab}(\alpha \theta)~.
\eeq

\end{document}